\documentclass[preprint,prd,tightenlines]{revtex4-1}
\usepackage[utf8]{inputenc}
\usepackage{amsthm}
\usepackage{amsmath}
\usepackage{tikz}
\usepackage{makecell}
\usepackage{multirow}
\usepackage{hyperref}
\usepackage[normalem]{ulem}
\usepackage[a4paper, margin=2.0cm]{geometry}
\usepackage{comment}
\usepackage{orcidlink}

\usepackage{natbib}
\usepackage{graphicx}
\usepackage{amssymb}
\usepackage{subfiles}
\usepackage{color}
\usepackage{lineno}
\usepackage{booktabs}
\usepackage{amsmath}
\usepackage{amsfonts}
\usepackage{xspace}
\usepackage{siunitx}
\usepackage{physics}

\def\Bsignz     {\ensuremath{B_{\mathit{CP}}}\xspace}
\def\Btagnz     {\ensuremath{B_{\text{tag}}}\xspace}
\def\Btag       {\ensuremath{B^0_{\text{tag}}}\xspace}
\def\Bbartag    {\ensuremath{\bar{B}^0_{\text{tag}}}\xspace}

\def\Bsig       {\ensuremath{B^0_{\text{sig}}}\xspace}

\def\sigmadt    {\ensuremath{\sigma_{\Delta t}}\xspace}

\def\dmd        {\ensuremath{\Delta m_d}\xspace}

\def\deltat     {\ensuremath{\Delta t}\xspace}

\def\deltae     {\ensuremath{\Delta E}\xspace}

\def\deltattru  {\ensuremath{\Delta t_{\text{true}}}\xspace}
\def\sigmadt    {\ensuremath{\sigma_{\Delta t}}\xspace}

\newcommand{\ddeltat}{{\text{d}\ensuremath{\Delta \tau}}\xspace}

\newcommand{\fours}{\ensuremath{\Upsilon(4S)}\xspace}
\newcommand{\acp}{\ensuremath{A_\mathit{CP}}\xspace}
\newcommand{\scp}{\ensuremath{S_\mathit{CP}}\xspace}

\newcommand{\bjpsiks}{\ensuremath{B^0\rightarrow J/\psi K_S^0}\xspace}
\newcommand{\bdstpi}{\ensuremath{B^0\rightarrow D^{(*)-}\pi^+}}

\newcommand{\bpjpsikp}{\ensuremath{B^+\rightarrow J/\psi K^+}\xspace}
\newcommand{\bpdzpi}{\ensuremath{B^+\rightarrow \bar D^{0}\pi^+}}

\newcommand{\sinbb}{\ensuremath{\sin2\phi_1}\xspace}
\newcommand{\Dt}{\ensuremath{\Delta t}\xspace}
\newcommand{\taubz}{\ensuremath{\tau_{B^0}}\xspace}

\newcommand{\Dmd}{\ensuremath{\Delta m_d}\xspace}

\def\Bz     {\ensuremath{B^0}\xspace}
\def\Bbar   {\ensuremath{\bar B}\xspace}%{\kern 0.18em\overline{\kern -0.18em B}{}\xspace}
\def\Bzb    {\ensuremath{\Bbar^0}\xspace}

\def\KS     {\ensuremath{K^0_{\scriptscriptstyle S}}\xspace}

\def\jpsi   {\ensuremath{{J\mskip -3mu/\mskip -2mu\psi\mskip 2mu}}\xspace}

\begin{document}

\vspace*{-3\baselineskip}
\resizebox{!}{3cm}{\includegraphics{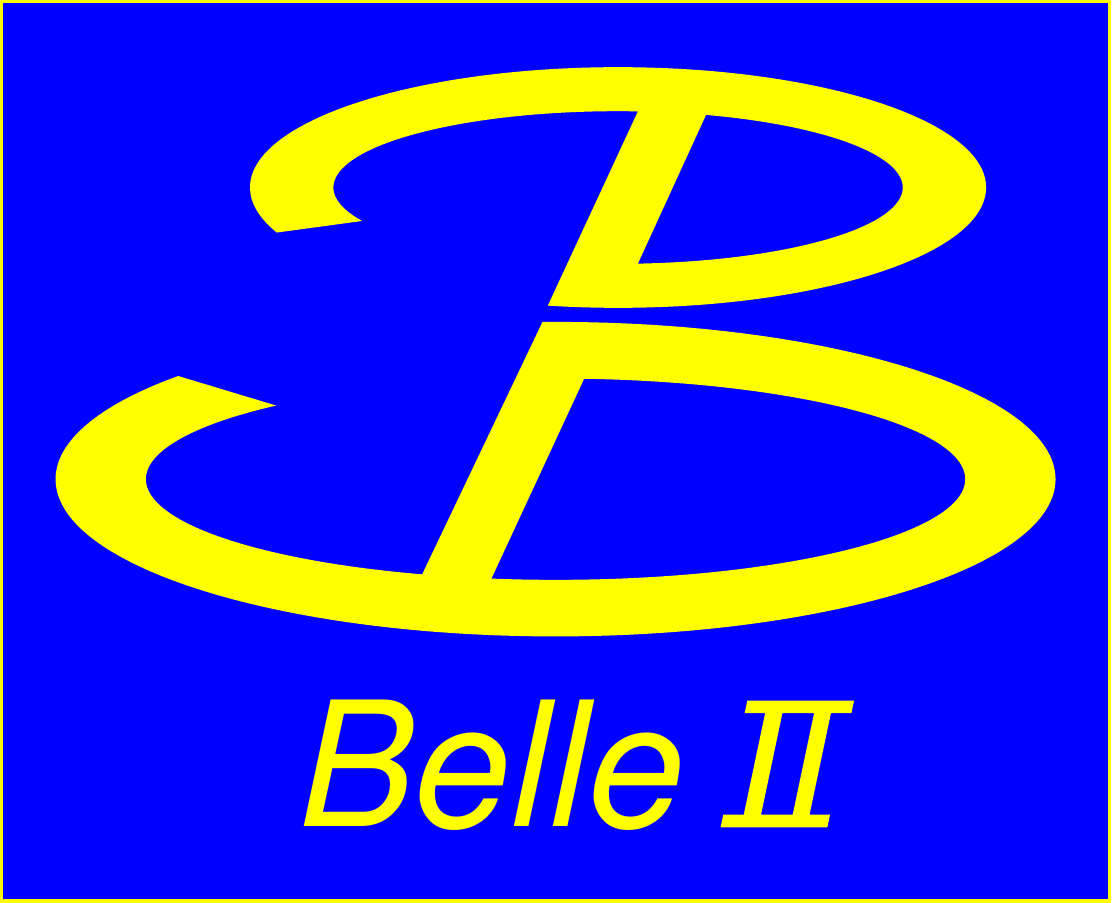}}

\vspace*{-5\baselineskip}
\begin{flushright}

February 24, 2023
\end{flushright}

\title{\quad\\[0.5cm] \boldmath Measurement of decay-time-dependent \textit{CP} violation  in $B^0 \to  \jpsi K^0_S$ decays using 2019-2021 Belle~II data}

%%% Paper:    (2022 conference papers)
%%% Journal:  (2022 conferences)
%%% Updates:
%%% November 17, 2022 - new iteration
%%% ====================================================================
%%% Use \input{authors-conf2022-orcid} to insert this material into your latex file.
  \author{F.~Abudin{\'e}n\,\orcidlink{0000-0002-6737-3528}} % 2250
  \author{I.~Adachi\,\orcidlink{0000-0003-2287-0173}} % 2590
  \author{K.~Adamczyk\,\orcidlink{0000-0001-6208-0876}} % 2239
  \author{L.~Aggarwal\,\orcidlink{0000-0002-0909-7537}} % 10083
  \author{P.~Ahlburg\,\orcidlink{0000-0002-9832-7604}} % 2367
  \author{H.~Ahmed\,\orcidlink{0000-0003-3976-7498}} % 11323
  \author{J.~K.~Ahn\,\orcidlink{0000-0002-5795-2243}} % 7423
  \author{H.~Aihara\,\orcidlink{0000-0002-1907-5964}} % 2223
  \author{N.~Akopov\,\orcidlink{0000-0002-4425-2096}} % 9443
  \author{A.~Aloisio\,\orcidlink{0000-0002-3883-6693}} % 2194
% ...
% 13403
\collaboration{The Belle II Collaboration}

\begin{abstract}
We report a measurement of the mixing-induced and direct \textit{CP} violation parameters 
\scp and \acp  from $\Bz\to \jpsi\KS$ decays reconstructed
 by the Belle~II experiment at the \mbox{SuperKEKB} asymmetric-energy electron-positron collider. The data, collected at the center-of-mass energy of the  $\Upsilon(4S)$ resonance, correspond to $190\;\text{fb}^{-1}$ of integrated luminosity.
We measure ${\scp = 0.720\pm0.062\pm0.016}$ and $\acp = 0.094\pm0.044^{+0.042}_{-0.017}$, where the first uncertainties are statistical and the 
second systematic. 
In the Standard Model, \scp equals $\sin(2\phi_1)$ to a good approximation.

\end{abstract}


{\let\newpage\relax\maketitle}

%%%%%%%%%%%%%%%%%%%%%%%%%%%%%%%%%%%%%%%%%%%%
% Theory & Intro
%%%%%%%%%%%%%%%%%%%%%%%%%%%%%%%%%%%%%%%%%%%%

\section{Introduction}

In the Standard Model (SM), \textit{CP} violation in the quark sector arises from a single irreducible phase in 
the Cabibbo-Kobayashi-Maskawa (CKM) quark-mixing matrix~\cite{Kobayashi:1973fv}.

Measuring decay-time-dependent \textit{CP}-violating rate asymmetries in neutral  $B$ meson decays to \textit{CP}~eigenstates
mediated by a $b\rightarrow c\bar{c} s$ 
tree-level transition gives access to  $\phi_1 \equiv \arg(-V^{}_{cd}V_{cb}^*/V^{}_{td}V_{tb}^*)$, 
one of the angles of the CKM unitarity triangle. 
The probability  of a $B$ meson having a flavor content $q$ ($q=-1$ for $B^0$ and $q=+1$ for $\bar{B}^0$) at some time and decaying to
 such a {\it CP}~eigenstate after a time \deltat is
\begin{equation}
    \label{eq:probdt}
    P(\deltat, q)=\frac{e^{-|\deltat|/\taubz}}{2\taubz}\{1+q[\scp\sin(\Dmd\Dt)  +\acp\cos(\Dmd\Dt)]\},
\end{equation}
where $\scp$ is the mixing-induced, and $\acp$ the direct, \textit{CP}-violating parameter; $\taubz$ is the $\Bz$ lifetime; 
and $\Dmd$ is the $\Bz$-$\Bzb$ oscillation frequency, which corresponds to the mass difference between the two 
neutral $B$ mass eigenstates.  
The SM predicts  $\acp=0$ and $\scp=-\eta\sinbb$, where $\eta$ is the \textit{CP} eigenvalue of the final state 
($\eta=+1$ for a \textit{CP}-odd and $\eta=-1$ for a \textit{CP}-even state).
The Belle, BaBar, and LHCb collaborations all measured \acp and \scp using
 several \textit{CP} eigenstates~\cite{Belle:2012paq,BaBar:2009byl,LHCb:2015ups,LHCb:2017mpa}. 
 All measurements have comparable precision, with the result of Belle being the most precise~\cite{Belle:2012paq},
\begin{equation*}
    \arraycolsep=0.4pt\def\arraystretch{1.5}
    \begin{array}{l r c l r c l}
        \scp = & 0.667  &  \pm & 0.023 \text{(stat)} & \pm & 0.012 \text{(syst)},  \\
        \acp = & 0.006  &  \pm & 0.016 \text{(stat)} & \pm & 0.012 \text{(syst)}.
    \end{array}
\end{equation*}

We present a Belle II measurement of these parameters using decays into  $B^0\to\jpsi\KS$  final states, which are \textit{CP}-odd, reconstructed from electron-positron collisions corresponding to $\SI{190}{fb}^{-1}$. 
The $B^0\to\jpsi\KS$  decay is the most sensitive channel for the \sinbb measurement due to its relatively
high branching fraction, 
low background, and small penguin pollution.  The latter  ensures that $\scp=\sinbb$ in this mode with an approximation better than $2\%$~\cite{Frings:2015eva}. 
As the data set is four times smaller than that used in the Belle analysis, the statistical precision is not yet competitive. 
The analysis, however, yields comparable systematic uncertainties, paving the way to a follow-up \sinbb measurement with 
competitive precision.

The $B$ mesons are produced in $e^+e^-\to\fours\to B\bar B$ events.
The SuperKEKB collider \cite{Ohnishi:2013fma} accelerates electron and positron beams to 7~GeV and 4~GeV, respectively,  producing \fours with a Lorentz boost $\beta\gamma=0.287$.
We fully reconstruct one $B$ meson, \Bsignz, in its decay to $\jpsi\KS$. The other $B$ meson, \Btag, is 
partially reconstructed by combining all charged particles not used in the \Bsignz reconstruction. We use a flavor-tagging
algorithm to determine the flavor of the \Btag at the time of its decay~\cite{Belle-II:2021zvj}. 
As the two $B$ mesons are in an entangled
quantum state, knowledge of the \Btag flavor determines the  \Bsignz flavor at that time. As the two $B$ mesons are almost at rest in the \fours frame, they are boosted 
in the lab frame and the two $B$ vertices are  displaced from each other. We measure their relative displacement 
to deduce the time difference \deltat between the \Bsignz and \Btagnz decays. We fit Eq.~(\ref{eq:probdt}) to the background-subtracted
\deltat distribution to measure \scp and \acp. Detection effects are controlled using  $B^0\rightarrow D^{(*)-}\pi^+$ decays.

%%%%%%%%%%%%%%%%%%%%%%%%%%%%%%%%%%%%%%%%%%%%%
%%%% Belle II detector
%%%%%%%%%%%%%%%%%%%%%%%%%%%%%%%%%%%%%%%%%%%%%
\section{Experimental setup}

The Belle~II detector  consists of several subsystems arranged in a cylindrical structure around the beam pipe~\cite{Belle-II:2010dht}.
The detector is asymmetric with more extensive coverage in the $e^-$~beam direction to mirror the energy asymmetry of the  beams.
The tracking system consists of a two-layer silicon-pixel detector (PXD) surrounded by a four-layer double-sided silicon-strip detector (SVD) and a 56-layer central drift chamber (CDC).
For the data used in this work, the second PXD layer is partially instrumented and covers only one sixth of the azimuthal angle.
In the case of \bjpsiks decays,    
the combined PXD and SVD system provides an  average \Bsignz vertex resolution along the beam  direction of approximately $\SI{25}{\micro m}$.
A time-of-propagation counter and an aerogel ring-imaging Cherenkov counter that cover the barrel and forward end-cap regions of the detector, respectively, are used for charged-particle identification.
An electromagnetic calorimeter fills the remaining volume inside a  superconducting  magnet that generates an axial, uniform 1.5~T field. 
It measures the energy of photons and supplements particle identification.
A dedicated system to identify $K^0_L$ mesons and muons is installed in the outermost part of the detector.
The $z$ axis of the laboratory frame is defined as the central axis of the solenoid, with its positive direction defined by the direction of the electron beam.
The data are processed using the Belle~II analysis software framework \cite{Kuhr:2018lps}, which relies on the track reconstruction algorithm described in Ref.~\cite{BelleIITrackingGroup:2020hpx}.

Simulated events are used for selection optimization, fit modelling, and validation of the measurement. 
They are generated using \texttt{KKMC} for quark-antiquark pairs from $e^+ e^-$ collisions~ \cite{Jadach:1999vf}, \texttt{PYTHIA8} for hadronization\cite{Sjostrand:2014zea}, \texttt{EVTGEN} for the decay of  hadrons~ \cite{Lange:2001uf}, and \texttt{GEANT4}  for the detector response~\cite{GEANT4:2002zbu}.
The simulation includes the effect of beam-induced background~\cite{Liptak:2021tog}.

%%%%%%%%%%%%%%%%%%%%%%%%%%%%%%%%%%%%%
% Event reconstruction
%%%%%%%%%%%%%%%%%%%%%%%%%%%%%%%%%%%%%

\section{Event Reconstruction}

We reconstruct the $\Bsignz$ decay by combining four charged particles.
Charged particle trajectories (tracks) are reconstructed with the PXD, SVD, and CDC.
All tracks are required to have a polar angle $\theta$ within the CDC acceptance, 
\textit{i.e.}, $17\si{\degree} <\theta<150\si{\degree}$.
Tracks that are not used to form a \KS candidate are required to have a distance-of-closest-approach 
to the interaction region of less than \SI{3}{cm} in the $z$ direction and less 
than \SI{0.5}{cm} in the transverse plane.

We require that electrons and muons be identified by their particle identification (PID) likelihoods, which
are constructed by combining information from several subdetectors.
The $\jpsi$ meson candidates are reconstructed from pairs of oppositely-charged electrons with an $e^+e^-$ mass 
in $[2.95,\;3.15]\;\si{GeV}/c^2$, and muons with a $\mu^+\mu^-$ mass 
in $[3.00,\;3.15]\;\si{GeV}/c^2$.
The $\KS$ meson candidates are reconstructed from pairs of oppositely-charged particles with a $\pi^+\pi^-$ mass 
in $[480,\;515]\;\si{MeV}/c^2$.
Candidate \KS mesons with a flight distance from the  $\jpsi$ decay vertex of less than $\SI{50}{\micro m}$ are rejected, in accordance with the long lifetime of the $\KS$.

The $\Bsignz$ meson candidates are reconstructed by combining  $\jpsi$ and $\KS$ candidates.
For each \Bsignz, we compute the beam-constrained mass $M_\mathrm{bc}$ and the energy discrepancy $\Delta E$. They are defined as  $M_{\rm bc}\equiv\sqrt{(E_{\rm beam}/c^2)^2-(p_B^*/c)^2}$ and  $\Delta E\equiv E_B^*-E_{\rm beam}$, where $E_{\rm beam} = \sqrt{s}/2$ is the energy of one beam in the center-of-mass frame, and $E_B^*$ ($p_B^*$) is the center-of-mass energy (momentum) of the  $\Bsignz$ candidate.
We require  $M_{\rm bc}>5.27~{\rm GeV}/c^2$ 
and $-0.10~{\rm GeV}<\Delta E<0.15~{\rm GeV}$.
The $J/\psi$ mass is constrained to its known value~\cite{PDG} in the  computation of \deltae to improve the resolution.
Events with the second to the zeroth Fox-Wolfram moments, $R_2$, larger than 0.4 are rejected to suppress $e^+e^-\to  q\bar{q}$ background, where $q$ is a $u$, $d$, $c$,  or $s$ quark~\cite{fw}.

We determine the \Btagnz vertex position and flavor using the remaining tracks in the event. Each is required to have at least one hit in each of the PXD, SVD, and CDC detectors and correspond to a particle of momentum greater than $\SI{50}{MeV}/c$. Each particle must also originate from the $e^+e^-$ interaction region according to the same criteria given above. We remove pairs of oppositely charged particles that have dipion mass compatible with the \KS mass.  

The $\Bsignz$ decay vertex is reconstructed using the {\tt TreeFitter} algorithm~\cite{Hulsbergen:2005pu,BelleII:2019dlq},
and the $\Btagnz$ decay vertex  using the {\tt Rave} algorithm~\cite{Waltenberger:2008zza}.
The latter mitigates the bias from tracks that are detached from the $\Btagnz$ decay vertex (typically coming from a charmed meson)
by assigning a small weight to tracks that yield a large contribution to the vertex $\chi^2$.
Each $B$ vertex is constrained to lie along a line having  the $e^+e^-$ 
interaction point as origin and the momentum of that $B$  as direction.

The proper-time difference $\Delta t\equiv t_\mathit{CP}-t_{\rm tag}$ is reconstructed as ${\Delta t \approx (l_\mathit{CP}-l_{\rm tag})/\beta\gamma \gamma^* c}$, where $l_\mathit{CP}$ and $l_{\rm tag}$ are the decay-vertex positions of $B_\mathit{CP}$ and $B_{\rm tag}$ projected onto the boost axis, $\beta\gamma$ is the $\Upsilon(4S)$ Lorentz boost factor   and $\gamma^*\approx 1.002$ is the Lorentz factor of the $B$ meson in the $\Upsilon(4S)$ frame. 
Additional quality criteria are applied to both the $\Btagnz$ and $\Bsignz$ vertex fits, and the  uncertainty on $\deltat$, $\sigmadt$, as determined
from uncertainties in track parameters, is required to be less than $2\;\text{ps}$.

The $b$ or $\bar b$ flavor of the \Btag meson is identified (``tagged'') using
inclusive properties of particles that are not associated with the reconstructed $\Bsignz \to \jpsi\KS$ candidate. 
Flavor tagging is performed using a category-based  algorithm~\cite{Belle-II:2021zvj}.
The flavor tagging information is represented by two parameters, the $b$-flavor charge $q$ and tag quality $r$.
The parameter $r$ %, also known as dilution factor, 
ranges from $r=0$ for no flavor discrimination to $r=1$ for unambiguous flavor assignment.

Events in which the fully reconstructed $B$ decays as \bdstpi\ (instead of \bjpsiks as for signal events)  are used to calibrate 
the flavor-tagger and \deltat resolution parameters. Their reconstruction and selection is 
described in Ref.~\cite{had}.
To validate the fit procedure, we determine the \textit{CP}-violation parameters using \textit{CP}-conserving $B^+ \to J/\psi K^+$ decays.
The hadronic $B^+$ decay mode $B^+ \to \bar{D}^0 \pi^+$, with $\bar{D}^0\to K^+\pi^-$ and $\bar{D}^0\to K^+\pi^- \pi^0$, is used to calibrate the resolution function
and flavor-tagger parameters for the charged decay modes.
The selection of the charged $B$ decay modes is kept as close as possible to that of the neutral $B$ modes. 
We require that the charged kaon in the $B^+ \to J/\psi K^+$ decay satisfy a  PID likelihood criterion. 
The requirement $R_2<0.4$ is used to  suppress the continuum background to \bpjpsikp and \bpdzpi decays.
The total reconstruction and selection efficiencies for \bjpsiks  and \bpjpsikp decay modes
are listed in Tab.~\ref{tab:result_table}.

After the selection, the fraction of events containing multiple candidate decays is approximately $0.4\%$ and $2\%$ for \bjpsiks and \bdstpi, respectively.
%, and approximately $2\%$ for \bdstpi.
All candidates are retained for further analysis. 

%%%%%%%%%%%%%%%%%%%%%%%%%%%%%%%%%%%%
% CP fits
%%%%%%%%%%%%%%%%%%%%%%%%%%%%%%%%%%%%

    \section{Determination of the sample composition}

\begin{figure}[h]
    \centering
    \includegraphics[width=0.68\linewidth]{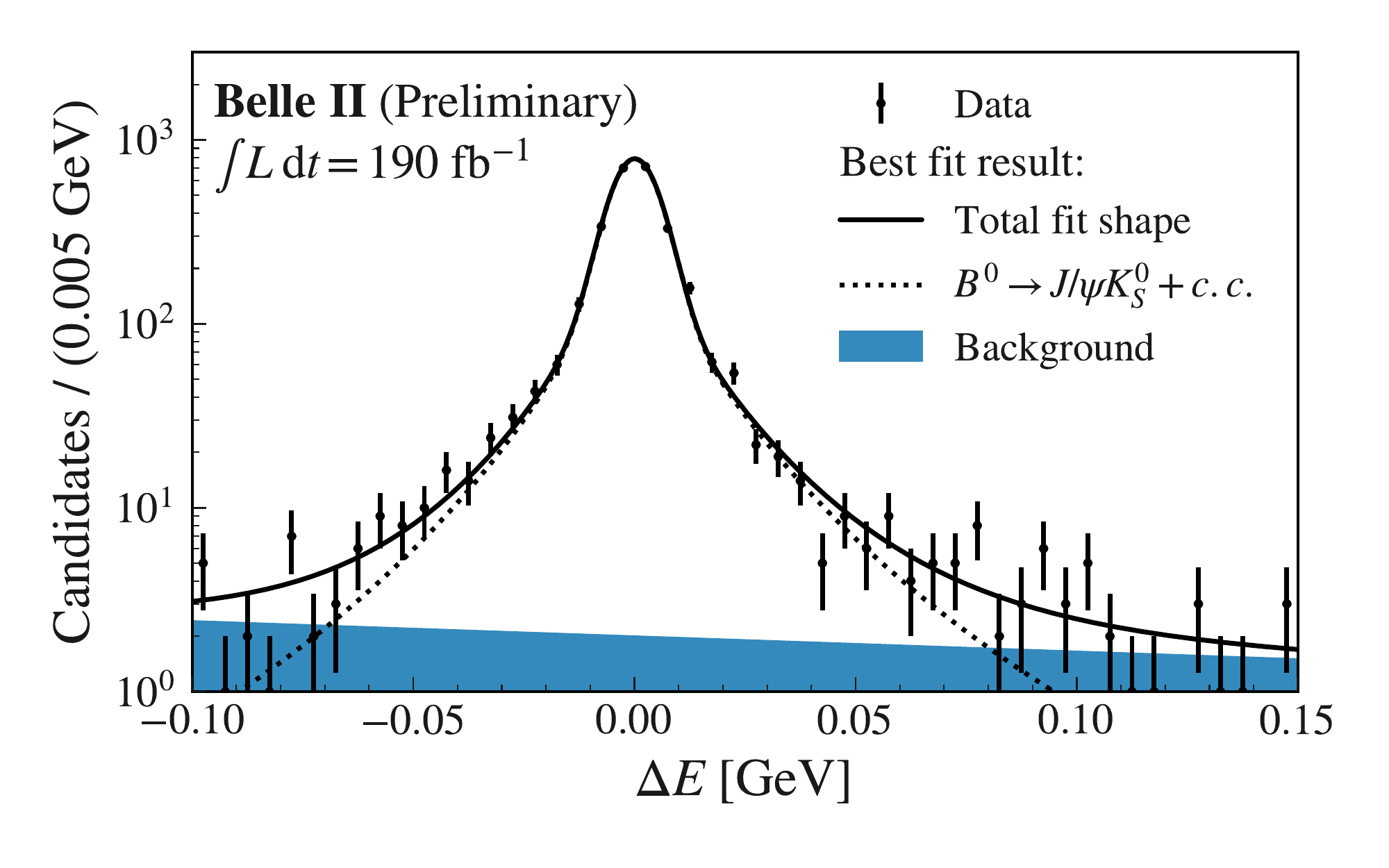}
    \caption{Distribution of \deltae for $B^0\to \jpsi(\to\mu^+\mu^-, e^+e^-)\KS$ candidates  reconstructed in the  data with the fit 
    curve superimposed. }
    \label{fig:defit_realdata}
\end{figure}

    We perform an unbinned extended maximum likelihood fit to the \deltae distribution to 
    measure the signal and background yields. 
    The signal PDF is modelled by the sum of a  Gaussian distribution and a double-sided Crystal Ball
    distribution~\cite{Gaiser:1982yw}.
    Dedicated studies using simulated events 
    show that
    the background consists of random combinations of tracks from  
    $e^+e^-\to q\bar q$
    events, or from $e^+e^-\rightarrow B\bar{B}$ or $e^+e^-\rightarrow B^+B^-$ events in which tracks from the two $B^{(\pm)}$ mesons are 
    combined. No significant background has a peak in \deltae. Therefore, the background shape is 
    modelled with an exponential distribution.  
    The observed $\Delta E$ distribution   is
    shown in Fig.~\ref{fig:defit_realdata} with the fitted curve superimposed. The signal yields and purities are listed in Tab.~\ref{tab:result_table}.
    Using event-by-event fractions calculated from the fitted \deltae PDFs, we employ the sWeight method~\cite{Pivk:2004ty} to obtain a background-subtracted \deltat
    distribution. We check that \deltae is not correlated to \deltat  in simulated \bjpsiks  signal and background events, hence allowing the use of sWeights.
    The sWeight method simplifies the analysis by avoiding the need to 
    parameterize the \deltat background distribution.

    \section{Calibration of detector effects}
    
    Three effects impact the measured \deltat distribution: the \deltat resolution, incorrect flavor-tagging assignment, and the small motion of the $B$ mesons in the \fours frame.
    These effects are calibrated by measuring the \deltat distribution for events in which the fully reconstructed $B$ decays as $B^0\to D^{(*)-}\pi^+$.
    
    Taking into account the asymmetry $\mu$ between the \Btag and \Bbartag reconstruction efficiencies, and the 
    fractions $w_{B}$ and $w_{\Bbar}$ of incorrectly tagged \Btag and \Bbartag mesons,
    the \deltat PDF in Eq.~(\ref{eq:probdt}) becomes
    \begin{equation}
        \begin{aligned}
        P_{\mathit{CP}}(\deltat,~q) 
         =\frac{e^{-|\Dt|/\taubz}}{2\taubz}&\{1-q\Delta w+q\mu(1-2w)\\
            &+[q(1-2w)+\mu(1-q \Delta w)][\scp\sin(\Dmd\Dt)  +\acp\cos(\Dmd\Dt)]  \},
        \end{aligned}
    \end{equation}
    where $w\equiv (w_{B}+w_{\Bbar})/2$ and $\Delta w\equiv w_{B}-w_{\Bbar}$. 

    The events are classified into  seven distinct intervals (bins) of $r$, with boundaries \linebreak $(0.0, 0.1, 0.25, 0.45, 0.6, 0.725, 0.875, 1.0)$. 
    The values of
    the flavor-tagger parameters are determined in each $r$ bin by performing an unbinned fit to  the background-subtracted \deltat distribution 
    of events in which the fully reconstructed $B$ decays as $B^0\to D^{(*)-}\pi^+$~\cite{had}. 
    As  the wrong-sign decay $\Bz\to D^{(*)+}\pi^-$ occurs with a negligible
    branching fraction, the charge of the $\pi$, $q_{\pi}$, identifies the flavor of the fully reconstructed $B$ at the time of its decay.
    For these events, the \deltat PDF is
     \begin{equation}
        \begin{aligned}
        P_{\text{flav}} (\deltat,~q,~q_{\pi}) 
         =\frac{e^{-|\Dt|/\taubz}}{4\taubz}&\{1-q\Delta w+q\mu(1-2w)\\
            &-q_{\pi}[q(1-2w)+\mu(1-q \Delta w)]\cos(\Dmd\Dt) \}.
        \end{aligned}
    \end{equation}
    
    To model the effect of the $\deltat$ resolution, the PDFs $P_{\text{flav}}$ and $P_{\mathit{CP}}$ are convolved with a resolution function using the parameterization of Ref.~\cite{had}.
    The resolution function is a sum of three components, referred to as core, tail, and outlier components, and is a function of the residual $\ddeltat = \deltat-\deltattru$.
    It reads
    \begin{equation}
        \begin{aligned}
            \mathcal{R}(\ddeltat;\sigmadt) &= f_{\text{core}}G(\ddeltat; m_{\text{core}}\sigmadt,s_{\text{core}}\sigmadt)\\ 
            &+f_{\text{tail}}\mathcal{R}_{\text{tail}}(\ddeltat; m_{\text{tail}}\sigmadt,s_{\text{tail}}\sigmadt, c/\sigmadt, f_{>}, f_{<})\\ &+f_{\text{OL}}G(\ddeltat; 0, \sigma_{\text{OL}}),
        \end{aligned}
    \end{equation}
    where $G(x; m, s)$ is a Gaussian distribution of mean $m$ and width $s$, and 
    $f_{\text{core}}+f_{\text{tail}}+f_{\text{OL}}=1$.
    The core  function accounts for approximately $70\%$ of events.
    The parameter $s_{\text{core}}$ is free in the fit to allow for an overall under- or over-estimate of the uncertainty \sigmadt. Its value is close to unity.
    The PDF $R_{\text{tail}}$ is parameterized as a sum of a Gaussian distribution and two exponentially modified Gaussian distributions,
    \begin{equation}
        \begin{aligned}
        \mathcal{R}_{\text{tail}}(x) &\propto G(x;m,s)\\
        &+f_{<} G(x;m,s)\otimes c\exp_{<}(cx)\\
        &+f_{>} G(x;m,s)\otimes c\exp_{>}(-cx),
        \end{aligned}
    \end{equation}
    where $\exp_{>}(-cx)=\exp(-cx)$ if $x>0$ and $\exp_{>}(-cx)=0$ otherwise, and
    similarly for $\exp_{<}(cx)$.
    The component $R_{\text{tail}}$ accounts for approximately $30\%$ of  events.
    The exponential tails, as well as $m_{\text{core}}$ and $m_{\text{tail}}$ that are allowed to take non-zero values, describe the impact of tracks originating from charmed mesons from the \Btagnz decay on the \Btagnz vertex resolution.
    The outlier component accounts for approximately $0.1\%$ of events that have poorly reconstructed vertices. It has a large width: $\sigma_{\text{OL}}=200\;\si{ps}$. 
    All resolution-function parameters share the same values across all $r$ bins, 
    apart from $m_{\text{core}}$, $m_{\text{tail}}$, and 
    $f_{\text{tail}}$ that have independent values in the seventh $r$ bin.
    This $r$ bin is mostly populated with \Btagnz mesons decaying semileptonically and is characterized by a resolution function with means closer to zero.
    Simulation studies show that neglecting this effect yields a bias in the measurement of \acp. 
    In the fit to the calibration modes, the following parameters are allowed to vary: $f_{\text{tail}}$, $m_{\text{core}}$, and $m_{\text{tail}}$ separately in $r$ bins $1$-$6$ and $7$, and $s_{\text{core}}$ and $s_{\text{tail}}$. The remaining parameters are fixed to their values derived from simulation studies. The corresponding systematic uncertainties are determined to be small.

    In addition to the convolution with the resolution function,
    the PDFs $P_{\mathit{CP}}$ and $P_{\text{flav}}$ are further corrected 
    for the effect of the small motion of the $B$ mesons in the \fours frame, 
    to yield the PDFs $P'_{CP}$ and $P'_{\text{flav}}$~\cite{had}.  
    The PDF $P'_{\text{flav}}$ is fitted to the background-subtracted \deltat distribution in the calibration sample with \dmd and \taubz fixed to their known values. This fit 
    has 28 free parameters, all related to the detector response: 21 flavor tagger
    parameters and 7 \deltat resolution parameters.

    \section{Determination of the \textit{CP} asymmetries}

    The parameters \scp and \acp are obtained by maximizing the logarithm of the weighed likelihood $L$,
    \begin{equation}
        \label{eq:pseudoll}
        \log L = \sum_{i} sW_i\; \log P'_{CP}({\Delta t_i, q_i| \acp, \scp}),
    \end{equation}
    where the sum runs over all events in the sample associated to an sWeight $sW_i$, 
    a flavor $q_i$, and a
    decay time difference $\Delta t_i$. 
    The \textit{CP} asymmetries 
    \acp and \scp are the only free parameters in the fit, \taubz and \dmd are fixed to the known values, and the resolution and flavor-tagging parameters are fixed to the values obtained with the calibration samples.
    We check that the fit procedure is unbiased by generating and fitting background-free simplified simulated experiments with various true values of \acp and \scp.

    The statistical uncertainties on the final result are computed by resampling, 
    \textit{i.e.}, bootstrapping~\cite{Efron:1979bxm}, the data of the calibration 
    and signal samples $1000$ times
    and computing the standard deviations of the fitted parameters across the 
    bootstrapped replicas. 
    For consistency checks, statistical uncertainties are estimated 
    using likelihood-ratio-based confidence regions applied to Eq.~\ref{eq:pseudoll}. The resulting uncertainties are typically
    underestimated by a few percent, as they do not account for the statistical fluctuations of the background, and of the flavor tagger and resolution-function parameters.

\begin{figure}[htbp]
        \centering
        \includegraphics[width=0.49\linewidth]{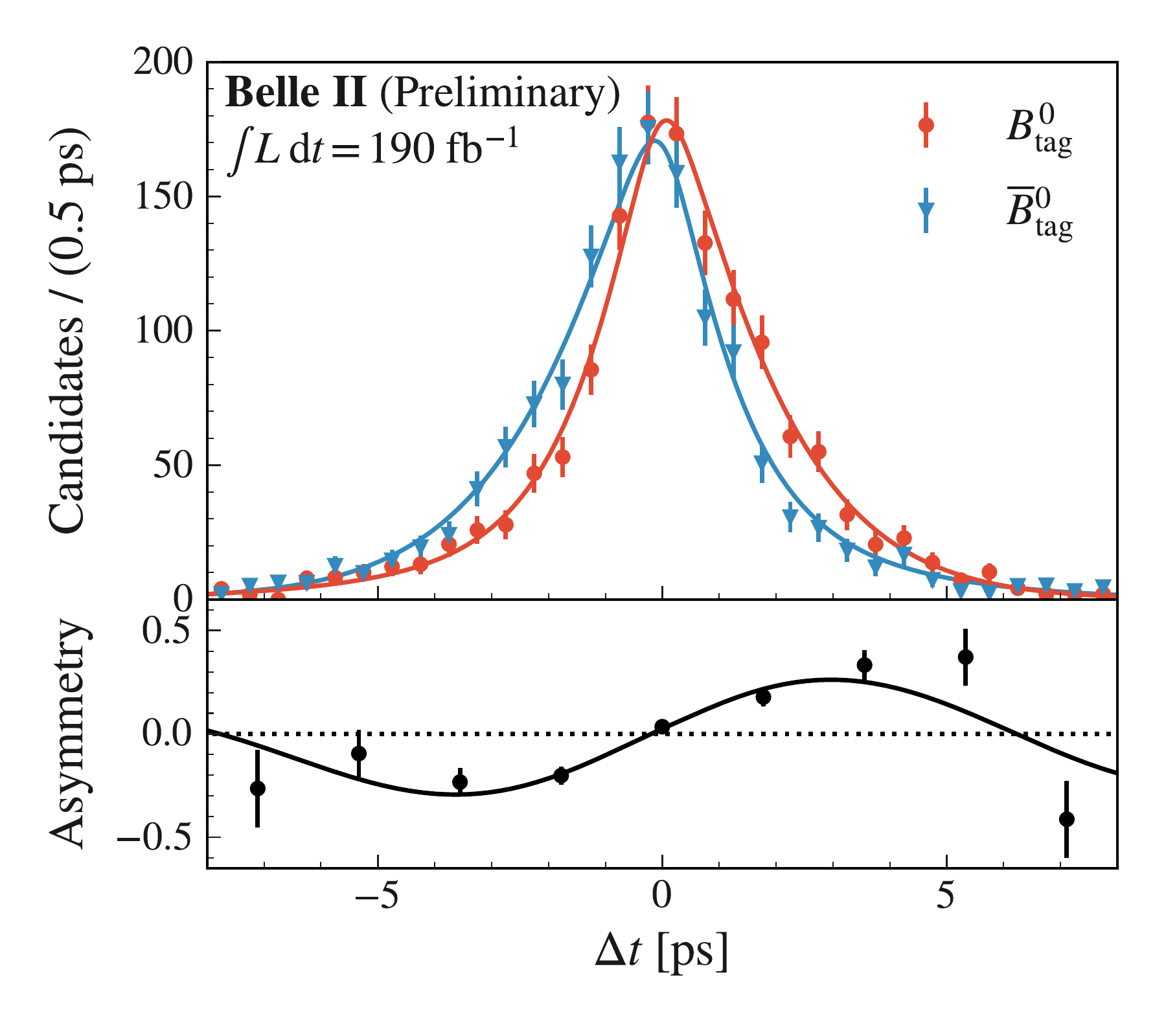}
        \includegraphics[width=0.49\linewidth]{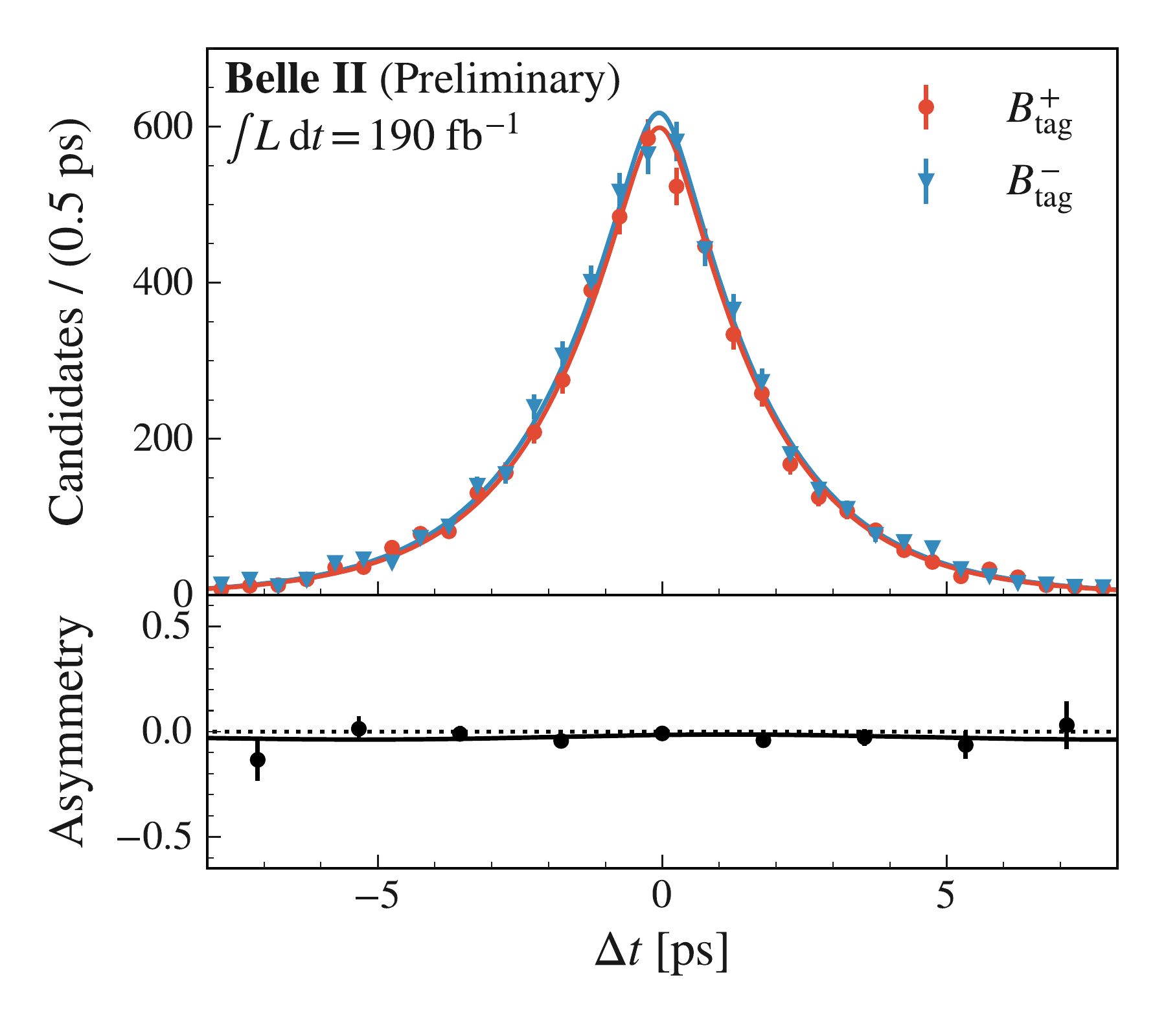}
        \caption{sWeighted \deltat distributions of  \bjpsiks  (left) and 
         $B^+\to\jpsi K^+$ (right) decays, separated by $\Btagnz$ flavor. 
        The fit projections are shown by solid curves and the asymmetry, defined as
        $(N({B}^0_{\text{tag}})-N(\bar{B}^0_{\text{tag}}))/(N({B}^0_{\text{tag}})+N(\bar{B}^0_{\text{tag}}))$ for the neutral $\Btagnz$ or 
        $(N({B}^+_{\text{tag}})-N({B}^-_{\text{tag}}))/(N({B}^+_{\text{tag}})+N({B}^-_{\text{tag}}))$ for the charged $\Btagnz$, is displayed underneath.}
        \label{fig:fitresult}
    \end{figure}

    The sWeighted \deltat distributions for the \bjpsiks sample are shown in Fig.~\ref{fig:fitresult} 
    separately for  $\bar{B}^0_{\text{tag}}$ and ${B}^0_{\text{tag}}$ events. The fit shapes, 
    corresponding to $P'_{\mathit{CP}}(\deltat,q=+1)$ and $P'_{\mathit{CP}}(\deltat,q=-1)$,
    are superimposed to the distributions and the raw asymmetry  is also shown.
    The fitted values of \scp and \acp are listed Tab.~\ref{tab:result_table}.
    The  values of \scp and \acp  obtained from the fit performed separately to the $\Bsignz \rightarrow J/\psi(\to e^+e^-) \KS$ and $\Bsignz \rightarrow J/\psi(\to\mu^+\mu^-) \KS$ subsamples are also listed.
    The uncertainties quoted in Tab.~\ref{tab:result_table} are statistical only and are estimated
    based on likelihood-ratios applied to Eq.~\ref{eq:pseudoll}.
    The fit results across the subsamples are in statistical agreement.

    As a check of our fitting procedure, we measure \acp and \scp for a sample of $B^+\rightarrow J/\psi K^+$  decays, which have no mixing-induced \textit{CP} violation and are expected to have negligible direct \textit{CP} violation. 
    For this consistency check, the $P'_{CP}$ PDF is fitted to the $B^+\rightarrow J/\psi K^+$ \deltat distribution, with the lifetime fixed to the known $B^+$ 
    lifetime and the oscillation frequency set to the value used for the neutral modes 
    to provide sensitivity to the \scp term. 
    %Similarly as for the neutral $B$ decays,
    All required parameters of the flavor tagger and resolution function are obtained by  fitting $P'_{\text{flav}}$ to the background-subtracted \deltat distribution of \bpdzpi decays.  
    The fitted values of \acp and \scp are shown in Tab.~\ref{tab:result_table} and are both consistent with zero.

    \begin{table}[htbp]
        \centering
        \caption{Total number of reconstructed events $N_{\text{evts}}$,  signal purity $p_{\text{sig}}$ in the signal region
        $|\deltae|<0.05\;\text{GeV}$, total reconstruction and selection efficiency 
        $\varepsilon_{\text{sig}}$ (including acceptance), and fit results for \acp and \scp
        in the  $\Bsignz\to\jpsi\KS$ and $B^+\rightarrow \jpsi K^+$ samples.
        Results are also shown separately for signal $B$ mesons reconstructed using  $J/\psi\rightarrow \mu^+\mu^-$ or
        $J/\psi\rightarrow e^+e^-$ decays. The uncertainties are statistical only.}
        \label{tab:result_table}
        \begin{tabular}{l r c c c c c}
            \hline\hline
            Sample & $N_{\text{evts}}$ & $p_{\text{sig}} (\%)$ & $\varepsilon_{\text{sig}}(\%)$ & \scp & \acp\\
            \hline
            \bjpsiks & 2755 & $98.6$ & $40.6$ &  ~$\phantom{-}0.720\pm0.062$~ & ~$0.094\pm0.044$~ \\% ~$\phantom{-}0.720\pm0.060$~ & ~$0.094\pm0.044$~ \\
            \hline
            $B^0\rightarrow \jpsi(\to\mu^+\mu^-) \KS$   & 1615 & $99.2$ &  $47.6$ & ~$\phantom{-}0.776\pm0.078$~ & ~$0.042\pm0.057$~ \\
            $B^0\rightarrow \jpsi(\to e^+e^-) \KS$ & 1140 & $98.0$ & $33.6$ & ~$\phantom{-}0.676\pm0.093$~ & ~$0.185\pm0.068$~ \\
            \hline
            \hline
            $B^+\rightarrow \jpsi K^+$ & 9973 & $98.1$ & $40.3$ & ~$\phantom{-}0.016\pm0.029$~ & ~$0.021\pm0.021$~ \\
            \hline
            $B^+\rightarrow \jpsi(\to \mu^+\mu^-) K^+$ & 5760 &  $99.0$ & $46.6$ & ~$-0.015\pm0.039$~ & ~$0.008\pm0.028$~ \\
            $B^+\rightarrow \jpsi(\to e^+e^-) K^+$ & 4213 & $96.7$ & $34.1$ & ~$\phantom{-}0.058\pm0.045$~ & ~$0.040\pm0.033$~ \\
            \hline\hline
        \end{tabular}
    \end{table}
    
    %To check for potential mis-modelling of the resolution 
    %function, 
    The $B^0$ lifetime and $B^0$-$\bar{B}^0$ oscillation-frequency measurements using \bdstpi decays in Ref.~\cite{had}  indicate that the resolution model used describes the data well. 
    The $B^0$ and $B^+$ lifetimes are measured using \bjpsiks, \bpdzpi,
    and $B^+\rightarrow \jpsi K^+$ decays, and are all found to be compatible with the known values. The values are $\tau_{B^0}=1.528 \pm 0.033\;\text{ps}$, $\tau_{B^+}=1.647\pm0.012\;\text{ps}$, 
    and $\tau_{B^+}=1.673 \pm 0.019\;\text{ps}$, respectively. The uncertainties are statistical only.
    To check for a potential bias that would originate from correlations between \deltae and \deltat,
    the data are divided into several subsamples corresponding to disjoint intervals of \deltat. 
    The \deltae fit is repeated and the sWeights computed in each subsample. The fit results are equal to those from the central fit,  indicating that potential correlations between \deltae and \deltat have a negligible impact on the final result.

%%%%%%%%%%%%%%%%%%%%%%%%%%%%%%%%%%%%
% Systematics
%%%%%%%%%%%%%%%%%%%%%%%%%%%%%%%%%%%%
\section{Systematic uncertainties}

 The statistical uncertainties computed
using bootstrap are reported in Tab.~\ref{tab:uncertainties}, together with the breakdown of the individual sources of systematic uncertainty.

\begin{table}[htbp]
    \centering
    \caption{Summary of the individual sources of uncertainties.}
    \label{tab:uncertainties}
    \begin{tabular}{l c c}
        \hline
        \hline
        {Source} & \unboldmath $\sigma(\scp)$ & \unboldmath $\sigma(\acp)$ \\ 
        \hline
         {Statistical} & \unboldmath ~0.0622~ & \unboldmath ~0.0439~ \\
         \hline
        \rule{0pt}{3ex}Calibration with \bdstpi decays\\
         ~~~\bdstpi  sample size                      & ~0.0111~ & ~0.0093~ \\
         ~~~Signal charge-asymmetry                   & ~0.0027~ & ~0.0126~ \\
         ~~~$w^+_6=0$ limit                           & ~0.0014~ & ~0.0001~ \\
         \rule{0pt}{3ex}Fit model\\
         ~~~Analysis bias                             & ~0.0080~ & ~0.0020~ \\
         ~~~Fixed resolution parameters               & ~0.0039~ & ~0.0008~ \\
         ~~~$\sigma_{\Delta t}$ binning               & ~0.0050~ & ~0.0051~  \\
         ~~~$\tau_{B^0}$, $\dmd$                      & ~0.0007~ & ~0.0002~ \\
         \rule{0pt}{3ex}\deltat measurement\\
         ~~~Alignment                                 & ~0.0020~ & ~0.0042~ \\
         ~~~Beam spot                                 & ~0.0024~ & ~0.0020~ \\
         ~~~Momentum scale                            & ~0.0005~ & ~0.0013~ \\
        \rule{0pt}{3ex}\bjpsiks \deltae background shape         & ~0.0037~ & ~0.0015~ \\
         Multiple candidates                          & ~0.0005~ & ~0.0008~\\
         \textit{CP} violation in \Btag decays      & ~0.0020~ & ~$^{+0.0380}_{-0.0000}$\\
         \hline
         {Total systematic}                 & \unboldmath $0.0163$ & \unboldmath $^{+0.0418}_{-0.0174}$\\
         \hline
         \hline
    \end{tabular}

\end{table}

Several systematic uncertainties are related to the calibration of the resolution function and flavor-tagger parameters with \bdstpi decays.
The statistical uncertainties due to the {size of the \bdstpi} sample are the dominant systematic uncertainty on \scp. The statistical uncertainties on 
the flavor-tagging parameters have a larger impact on the precision  than the statistical uncertainties on the resolution-function parameters. 
Another source of systematic uncertainty related to the calibration of detector effects is the {signal charge-asymmetry}, reflecting the fact that the reconstruction and selection efficiencies for $B^0\to D^{(*)-} \pi^+$ and the charge-conjugated $\bar{B}^0\to D^{(*)+} \pi^-$ channels differ by $2.6\%$.
In the nominal configuration of the fit, these efficiencies are assumed to be equal, which induces 
a bias on the \Btag detection asymmetries $\mu$.
To evaluate the impact on \scp and \acp, 
the measurement of the flavor-tagging parameters is repeated after weighing the \bdstpi data such that the event yields for $B^0$ and $\bar{B}^0$ decays are identical. The difference between the \scp and \acp values
obtained with and without this correction is assigned as a systematic uncertainty.
In the \deltat fit to the calibration modes, the wrong-tag fraction $w_{B^0}$ in the highest $r$ bin is evaluated to be at the lower limit $w_{B^0} = 0$ in $30\%$ of the resampled (bootstrapped) replicas obtained from the data. 
To estimate the possible bias, we perform an alternative fit where $w_{B^0}$ is not determined by the fit, but fixed to $0.0$, and the difference of $\scp$ and $\acp$ with respect to the values from the nominal fit is assigned as the uncertainty.

Other systematic uncertainties are related to the fit model, \textit{i.e.}, inaccuracies in the parameterization of $P'_{CP}$ and $P'_{\text{flav}}$.
The {analysis bias} is the difference between the true parameter values used in the
simulation and the value obtained by fitting $500$ simulated samples with an average number of events corresponding to that observed in the experimental data.
The systematic uncertainty related to the  {resolution parameters} fixed to the their values in the simulation is evaluated by repeating the fit, while letting the otherwise  fixed  parameters free to vary one-by-one. Each fit yields additional values for \acp 
and \scp whose deviations from the nominal values are
summed in quadrature to yield a systematic uncertainty.  
In the implementation of the fitter,  \sigmadt is used as a conditional
observable. The \sigmadt distributions are sampled in each of the $r$ bins and separately for $B^0$ and $\bar{B}^0$ events.
To assess the  uncertainty related to the choice of {\sigmadt\ binning}, the number of bins in the \sigmadt  histograms is varied from 30 to 1000 bins.
The largest variation with respect to the nominal fit is assigned as a systematic uncertainty.
The impact of the uncertainty of the $B^0$ lifetime and the $B^0$-$\bar{B}^0$ oscillation frequency is tested by varying these parameters within their uncertainties~\cite{PDG}.

Another category of systematic uncertainties are those related to the measurement of \deltat.
To evaluate the systematic uncertainty related to the detector alignment~\cite{Bilka:2021rqj}, simulated \bdstpi and \bjpsiks events are reconstructed with four alternative misalignment scenarios.
The extraction of \scp and \acp, including the calibration with the \bdstpi modes, is repeated using the four misaligned samples,
and the maximal deviation from the nominal configuration is assigned as {alignment} uncertainty.
Both \Bsig and \Btag decay vertices are determined with the $\Upsilon(4S)$ production vertex constrained to the {beam spot} position.
The beam spot position, dimensions, and orientation are continuously monitored using  $e^+e^- \to \mu^+\mu^-$ events.
The uncertainties on these parameters originate from the limited size of the $e^+e^- \to \mu^+\mu^-$ data sample and the systematic uncertainty of the measurement.
To estimate the impact on $\scp$ and $\acp$, the analysis is repeated with the beam-spot parameters shifted by their respective uncertainties.
The difference between the {momentum scale} of tracks in the data and in the simulation is measured to be less than $0.1\%$ and has a very small impact on the analysis.

In addition to the sources of uncertainty mentioned above, several other systematic effects are considered. 
The measurement is repeated by changing the parameterization of the {\deltae background shape} from an exponential  to a second-degree polynomial.   The deviation with respect to the nominal is assigned as a systematic uncertainty. 
To assess a possible bias stemming from {multiple candidates}, 
the fit is repeated by removing multiple candidates at random.
The  difference of this fit result with respect to the nominal is taken as a systematic uncertainty.
The expression of $P'_{CP}$ does not account for
the effect of {\textit{CP} violation in \Btag decays}~\cite{Long:2003wq}. 
This yields a systematic uncertainty determined following Ref.~\cite{Belle:2012paq}. This is the dominant source of systematic uncertainty on \acp.

The total systematic uncertainty is evaluated as the quadratic sum of the individual contributions.
The precision on \scp and \acp is limited by the sample size. The systematic uncertainty of $0.016$ on $\scp$  is one fourth of the statistical uncertainty and is comparable to the world-average precision.

%%%%%%%%%%%%%%%%%%%%%%%%%%%%%%%%%%%%
% Conclusion
%%%%%%%%%%%%%%%%%%%%%%%%%%%%%%%%%%%%
\section{Results and conclusions}
A measurement of   mixing-induced and direct \textit{CP} violation %parameters \scp and \acp 
in $B^0\rightarrow\jpsi \KS$  decays is performed using data collected by the Belle~II detector.
We find 2755 signal candidates in a sample consisting of $200\times 10^6$ $B\bar{B}$ pairs, where both $\jpsi \to \mu^+\mu^-$ and $\jpsi \to e^+ e^-$ decay channels are reconstructed.
The results are
\begin{equation*}
    \arraycolsep=0.4pt\def\arraystretch{1.5}
    \begin{array}{l r c l r c l}
        \scp = & 0.720  &  \pm & 0.062 \text{(stat)} & \pm & 0.016 \text{(syst)},  \\
        \acp = & 0.094  &  \pm & 0.044 \text{(stat)} & ^{\displaystyle{+}}_{\displaystyle{-}} & ^{\displaystyle{0.042}}_{\displaystyle{0.017}} \text{(syst)},
        %\acp = & 0.094  &  \pm & 0.044 \text{(stat)} & ^{\displaystyle{+}}_{\displaystyle{-}} & ^{{0.042}}_{{0.017}} \text{(syst)}.
    \end{array}
\end{equation*}
with a statistical correlation coefficient of $-6\%$. These results allow the determination of the CKM angle $\phi_1$~\cite{Frings:2015eva}; for negligible penguin pollution, as expected for this final state, our value for \scp corresponds to $\phi_1=(23.0\pm2.6\text{(stat)}\pm0.7\text{(syst)})\si{\degree}$.

These results are consistent with the world-average results. The statistical uncertainty is twice that of the current most precise determination, consistent with a four-times smaller data set. The systematic uncertainties are comparable. 

\begin{acknowledgments}
% Policy from October 20, 2022
This work, based on data collected using the Belle II detector, which was built and commissioned prior to March 2019, was supported by
%Armenia
Science Committee of the Republic of Armenia Grant No.~20TTCG-1C010;
%Australia
Australian Research Council and research Grants
No.~DE220100462,
No.~DP180102629,
No.~DP170102389,
No.~DP170102204,
No.~DP150103061,
No.~FT130100303,
No.~FT130100018,
and
No.~FT120100745;
%Austria
Austrian Federal Ministry of Education, Science and Research,
Austrian Science Fund
No.~P~31361-N36
and
No.~J4625-N,
and
Horizon 2020 ERC Starting Grant No.~947006 ``InterLeptons'';
%Canada
Natural Sciences and Engineering Research Council of Canada, Compute Canada and CANARIE;
%China
Chinese Academy of Sciences and research Grant No.~QYZDJ-SSW-SLH011,
National Natural Science Foundation of China and research Grants
No.~11521505,
No.~11575017,
No.~11675166,
No.~11761141009,
No.~11705209,
and
No.~11975076,
LiaoNing Revitalization Talents Program under Contract No.~XLYC1807135,
Shanghai Pujiang Program under Grant No.~18PJ1401000,
Shandong Provincial Natural Science Foundation Project~ZR2022JQ02,
and the CAS Center for Excellence in Particle Physics (CCEPP);
%Czech Republic
the Ministry of Education, Youth, and Sports of the Czech Republic under Contract No.~LTT17020 and
Charles University Grant No.~SVV 260448 and
the Czech Science Foundation Grant No.~22-18469S;
%EU
European Research Council, Seventh Framework PIEF-GA-2013-622527,
Horizon 2020 ERC-Advanced Grants No.~267104 and No.~884719,
Horizon 2020 ERC-Consolidator Grant No.~819127,
Horizon 2020 Marie Sklodowska-Curie Grant Agreement No.~700525 "NIOBE"
and
No.~101026516,
and
Horizon 2020 Marie Sklodowska-Curie RISE project JENNIFER2 Grant Agreement No.~822070 (European grants);
%France
L'Institut National de Physique Nucl\'{e}aire et de Physique des Particules (IN2P3) du CNRS (France);
%Germany
BMBF, DFG, HGF, MPG, and AvH Foundation (Germany);
%India
Department of Atomic Energy under Project Identification No.~RTI 4002 and Department of Science and Technology (India);
%Israel
Israel Science Foundation Grant No.~2476/17,
U.S.-Israel Binational Science Foundation Grant No.~2016113, and
Israel Ministry of Science Grant No.~3-16543;
%Italy
Istituto Nazionale di Fisica Nucleare and the research grants BELLE2;
%Japan
Japan Society for the Promotion of Science, Grant-in-Aid for Scientific Research Grants
No.~16H03968,
No.~16H03993,
No.~16H06492,
No.~16K05323,
No.~17H01133,
No.~17H05405,
No.~18K03621,
No.~18H03710,
No.~18H05226,
No.~19H00682, % Niigata
No.~22H00144,
No.~26220706,
and
No.~26400255,
the National Institute of Informatics, and Science Information NETwork 5 (SINET5), 
and
the Ministry of Education, Culture, Sports, Science, and Technology (MEXT) of Japan;  
%Korea
National Research Foundation (NRF) of Korea Grants
No.~2016R1\-D1A1B\-02012900,
No.~2018R1\-A2B\-3003643,
No.~2018R1\-A6A1A\-06024970,
No.~2018R1\-D1A1B\-07047294,
No.~2019R1\-I1A3A\-01058933,
No.~2022R1\-A2C\-1003993,
and
No.~RS-2022-00197659,
Radiation Science Research Institute,
Foreign Large-size Research Facility Application Supporting project,
the Global Science Experimental Data Hub Center of the Korea Institute of Science and Technology Information
and
KREONET/GLORIAD;
%Malaysia
Universiti Malaya RU grant, Akademi Sains Malaysia, and Ministry of Education Malaysia;
%Mexico
% CINVESTAV-IPN, UNAM, UAS, BUAP and CONACYT are funded under
Frontiers of Science Program Contracts
No.~FOINS-296,
No.~CB-221329,
No.~CB-236394,
No.~CB-254409,
and
No.~CB-180023, and No.~SEP-CINVESTAV research Grant No.~237 (Mexico);
%Poland
the Polish Ministry of Science and Higher Education and the National Science Center;
%Russia
the Ministry of Science and Higher Education of the Russian Federation,
Agreement No.~14.W03.31.0026, and
the HSE University Basic Research Program, Moscow;
%Saudi Arabia
University of Tabuk research Grants
No.~S-0256-1438 and No.~S-0280-1439 (Saudi Arabia);
%Slovenia
Slovenian Research Agency and research Grants
No.~J1-9124
and
No.~P1-0135;
%Spain
Agencia Estatal de Investigacion, Spain
Grant No.~RYC2020-029875-I
and
Generalitat Valenciana, Spain
Grant No.~CIDEGENT/2018/020
%Taiwan
Ministry of Science and Technology and research Grants
No.~MOST106-2112-M-002-005-MY3
and
No.~MOST107-2119-M-002-035-MY3,
and the Ministry of Education (Taiwan);
%Thailand
Thailand Center of Excellence in Physics;
%Turkey
TUBITAK ULAKBIM (Turkey);
%Ukraine
National Research Foundation of Ukraine, project No.~2020.02/0257,
and
Ministry of Education and Science of Ukraine;
%USA
the U.S. National Science Foundation and research Grants
No.~PHY-1913789 % Indiana CEEM
and
No.~PHY-2111604, % Luther
and the U.S. Department of Energy and research Awards
No.~DE-AC06-76RLO1830, % PNNL
No.~DE-SC0007983, % Wayne State
No.~DE-SC0009824, % Florida
No.~DE-SC0009973, % VPI
No.~DE-SC0010007, % Duke
No.~DE-SC0010073, % South Carolina
No.~DE-SC0010118, % Carnegie Mellon
No.~DE-SC0010504, % Hawaii
No.~DE-SC0011784, % Cincinnati
No.~DE-SC0012704, % BNL
No.~DE-SC0019230, % Duke
No.~DE-SC0021274, % Mississippi
No.~DE-SC0022350; % Louisville
%last group
and
%Vietnam
the Vietnam Academy of Science and Technology (VAST) under Grant No.~DL0000.05/21-23.

% Policy from October 20, 2022
These acknowledgements are not to be interpreted as an endorsement of any statement made
by any of our institutes, funding agencies, governments, or their representatives.

We thank the SuperKEKB team for delivering high-luminosity collisions;
the KEK cryogenics group for the efficient operation of the detector solenoid magnet;
the KEK computer group and the NII for on-site computing support and SINET6 network support;
and the raw-data centers at BNL, DESY, GridKa, IN2P3, INFN, and the University of Victoria for offsite computing support.

\end{acknowledgments}

\bibliographystyle{belle2-note}
\bibliography{references}

\end{document}